\documentclass[english,twocolumn,showpacs,superscriptaddress]{revtex4}
\usepackage[T1]{fontenc}
\usepackage[latin1]{inputenc}
\usepackage{graphicx}
\usepackage{amssymb,amsmath}

\newcommand{\bk}{\mathbf{k}}
\newcommand{\bq}{\mathbf{q}}
\newcommand{\bp}{\mathbf{p}}
\newcommand{\bu}{\mathbf{u}}
\newcommand{\bb}{\mathbf{b}}
\newcommand{\bn}{\mathbf{n}}
\newcommand{\bx}{\mathbf{x}}
\newcommand{\by}{\mathbf{y}}

\newcommand{\myvector}[1]{\mathbf{#1}}
\newcommand{\vU}{\myvector{U}}
\newcommand{\vB}{\myvector{B}}
\newcommand{\vX}{\myvector{X}}
\newcommand{\vY}{\myvector{Y}}
\newcommand{\vO}{\myvector{O}}
\newcommand{\vA}{\myvector{A}}
\newcommand{\vQ}{\myvector{Q}}
\newcommand{\vW}{\myvector{W}}
\newcommand{\vD}{\myvector{D}}

\newcommand{\prodscal}[2]{\langle#1|#2\rangle}

\makeatletter

\usepackage{psfrag}

\usepackage{babel}
\makeatother
\begin{document}

\title{Energy transfers in shell models for MHD turbulence}

\author{Thomas Lessinnes}

\affiliation{Physique Statistique et Plasmas, CP231, Facult\'e des Sciences, Universit\'e Libre de Bruxelles, B-1050
Bruxelles, Belgium}

\author{Mahendra K. Verma}

\affiliation{Department of Physics, Indian Institute of Technology Kanpur, India 208016}

\author{Daniele Carati}

\affiliation{Physique Statistique et Plasmas, CP231, Facult\'e des Sciences, Universit\'e Libre de Bruxelles, B-1050
Bruxelles, Belgium}

\begin{abstract}

A systematic procedure to derive shell models for MHD turbulence is proposed. It takes into account the conservation of ideal quadratic invariants such as the total energy, the cross-helicity and the magnetic helicity as well as the conservation of the magnetic energy by the advection term in the induction equation. This approach also leads to simple expressions for the energy exchanges as well as to unambiguous definitions for the energy fluxes. When applied to the existing shell models with nonlinear interactions limited to the nearest neighbour shells, this procedure reproduces well known models but suggests a reinterpretation of the energy fluxes.

\end{abstract}


\maketitle


\section{Introduction}

Understanding the existence and the dynamics of the magnetic field of the Earth, of the Sun and, in general, of other celestial bodies remains one of the most challenging problems of classical physics. Astronomical and geophysical observations have provided many insights into these phenomena~\cite{Moffat, Krause, Bran}. Laboratory experiments~\cite{Monch,Fauv} have confirmed that generation of magnetic field (dynamo) can take place under various circumstances, and lead to a variety of complex behaviours. However, analytical approaches of this problem are extremely complicated while numerical efforts are limited to a range of parameter space that is often quite distant from the realistic systems. For instance, in certain astrophysical bodies as well as in laboratory experiments, the kinematic viscosity $\nu$ of the fluid is six orders of magnitude smaller than its resistivity $\eta$. The two dissipation processes therefore takes place at very different time scales. This property makes direct numerical simulation of dynamo intractable.  Due to this reason we resort to simplified models.

Shell models specifically belong to this class of simplified approaches~\cite{bif03}. They have been constructed to describe interactions among various scales without any reference to the geometric structure of the problem. They were first introduced for fluid turbulence with the quite successful GOY shell model~\cite{Gletzer, YamadaOhkitani} and have been extended to MHD turbulence~\cite{Lrt, Frick, Gilbert, PS1}. In shell models, drastically reduced degrees of freedom (usually only one complex number) are used to describe the entire information provided by a shell of Fourier modes in wavenumber space. This approach reduces the description of turbulence from a partial differential equation to a reduced set of ordinary differential equations and provides a simplified tool for studying the energy and helicity exchanges between different scales at a significantly reduced numerical cost. 

The present work aims at deriving the expressions for the energy fluxes and the energy exchanges for magnetohydrodynamic (MHD) turbulence in a systematic and consistent manner.  Then we apply this scheme to study energy transfers in  a shell model of MHD.   This approach follows quite closely the previous efforts in which fluxes and energy exchanges have been identified for the complete MHD equation. However, in these works~\cite{davees01,ver04}, energy exchanges between two degrees of freedom have been determined from the triadic interactions up to an indeterminate circulating energy transfer. The strategy adopted in the present paper is somewhat different. Here, we derive the energy transfer formulas from the energy equations by identifying the terms that participate in these transfers.   This process also involves various symmetries and conservation laws of the ideal (dissipationless) equations.  For example, the energy transfer from a magnetic field shell to another magnetic field shell is dictated by the convective term of the induction equation that conserves the total magnetic energy. One of the main advantages of the present formalism is that we need not worry  about the indeterminate circulating transfer appearing in the related past work by Dar et al.~\cite{davees01,ver04}.   

The dynamo process involves growth of magnetic energy that is supplied from the kinetic energy by the nonlinear interactions.  As we will show in the paper, a clear and unambiguous identification of the various energy fluxes and energy exchanges between the velocity and the magnetic fields is very important in the study of dynamo effects. This is one of the main motivations for the development of the present approach.  The approach is also explicitly applied in Section~\ref{PS} to the derivation of the GOY shell model to MHD~\cite{PS1}. 

An outline of the paper is as follows:  A general formalism for expressing the various constraints satisfied by the nonlinearities in the shell models is discussed in Section~\ref{GenShel}. It is shown in Section~\ref{TF} that this formalism can be adapted nicely to the derivation of explicit expressions for the  energy fluxes as well as for the shell-to-shell energy exchanges in shell model.  In Section~\ref{PS}, we apply the formalism to the GOY shell model for MHD turbulence~\cite{PS1}, and study the energy energy fluxes for MHD turbulence.  In Section~V, we present our conclusions.


\section{Shell Models of MHD turbulence \label{GenShel} }

Shell models were first introduced for fluid turbulence (see for example~\cite{Gletzer,YamadaOhkitani,Sabra}). They can be seen as a drastic simplification of the Navier-Stokes or the MHD equations which, assuming periodic boundary conditions, are expressed in Fourier space as follows:
\begin{align}
\frac{d \bu_\bk}{dt}&=\bn_\bk(\bu,\bu)-\bn_\bk(\bb,\bb)-\nu k^2 \bu_\bk+\mathbf{f}_\bk\,,
\label{MHDequ}\\
\frac{d \bb_\bk}{dt}&=\bn_\bk(\bu,\bb)-\bn_\bk(\bb,\bu)-\eta k^2 \bb_\bk\,.\label{MHDeqb}
\end{align}
where $\bu_\bk$ and $\bb_\bk$ are the velocity and magnetic field Fourier modes respectively with wave vector $\bk$. The norm of this wave vector is  $k=|\bk|$. The viscosity $\nu$ and the magnetic diffusivity $\eta$ are responsible for the dissipative effects in these equations while energy is  injected through the forcing term $\mathbf{f}_\bk$. The nonlinear term is defined by
\begin{equation}
\bn_\bk(\bx,\by)=i \, \mathbf{P}(\bk)\cdot\sum_{\bp+\bq=-\bk} (\bk\cdot\bx^*_\bq) \by^*_\bp \,, \label{nonlingen}
\end{equation}
where $\bx$ and $\by$ can be either the velocity or the magnetic field. The tensor $\mathbf{P}$ is defined as
\begin{equation}
P_{ij}(\bk)=\frac{k^2\delta_{ij}-k_i\,k_j}{k^2}\label{tensorP}\,.
\end{equation}
It projects any field to its divergence-free part and it is used since only incompressible flows are considered in this study ($\nabla \cdot \bu=0$). In the velocity equation, the projection of the nonlinear terms using the tensor~(\ref{tensorP}) replaces the introduction of the pressure term. In the magnetic field equation, the nonlinear terms are usually not projected to their divergence-free parts. Indeed, the non divergence-free parts of the two nonlinear terms cancel each other and the constraint $\nabla \cdot \bb=0$ is automatically satisfied. The writing of the nonlinear term in the magnetic field equation using the form~(\ref{nonlingen}) has been used to stress and explore  the inner symmetries in the MHD equations.

The incompressible MHD equations are known to conserve the total energy, the cross helicity and the magnetic helicity.  The conservation of these quantities  plays a central role  for the derivation of shell models. Similarly, the conservation of both the kinetic helicity and the kinetic energy in absence of magnetic field are used to simplify further the shell model for MHD. There is however another property that has not been exploited so far: The conservation of magnetic energy by the first nonlinear term in the magnetic field equation. Indeed, assuming periodic boundary conditions, it is easy to prove that 
\begin{equation}
\sum_{\bk}\ \bn_\bk(\bu,\bb)\cdot\bb_\bk^* =0\,. \label{enerbcons}
\end{equation}   
The identification of a similar term in shell models for MHD will prove to be very useful in determining the energy exchanges and the energy fluxes in the shell model.

The equations of the evolution of the variables in a shell model are designed  to mimic as much as possible the MHD equations~(\ref{MHDequ}-\ref{MHDeqb}). In order to build the shell model using a systematic procedure, we first introduce the partition of the Fourier space into shells $s_i$ defined as the regions $|\bk|\in [k_{i-1},k_i]$ where $k_i=k_0\ \lambda^i$. In this definition, $k_0$ corresponds to the smallest wave vector. The number of shells is denoted by $N$, so that the wave vectors larger than $k_0\ \lambda^{N-1}$ are not included in the model. Any observable that would be represented in the original MHD equation by its Fourier modes $\mathbf{x}_\bk$ is described in the framework of the shell model by a vector of complex numbers noted $\vX$. Each component $x_i$ of this vector summarises the information from all the modes $\mathbf{x}_\bk$ corresponding to the shell $s_i$. It is also very useful to introduce the vector $\vX_i$ for which all components but the $i$-th are zero:
\begin{align}
\vX&=(x_1,x_2,...,x_N) \in \mathbb C^N\,,\\
\vX_i&=(0,0,...,0,x_i,0,...,0) \in \mathbb C^N\,,\\
\vX&=\sum_{i=1}^N \vX_i\,, \label{vecexpan}
\end{align}
where the expansion~(\ref{vecexpan}) is a direct consequence of the definition of $\vX_i$. 

In the following, the scalar product of two real fields will be needed for defining various quantities like kinetic and magnetic energies, cross helicity, kinetic and magnetic helicities. Using the Parseval's identity, the shell model version of this physical space scalar product is expressed as follows:
\begin{equation}
\prodscal{\vX}{\vY}\equiv \sum_{i=1}^N \frac{1}{2} (x_i \,  y_i^* +  y_i \,  x_i^*) \,.
\end{equation}
Due to the nonlinear evolution of the velocity and the magnetic field in the MHD equations, any attempt to design a mathematical procedure that would reduce the description of these fields to two vectors of complex numbers $\vU$ and $\vB$ must lead to closure issues. In the derivation of a shell model, the shell variables are usually not seen as projected versions of the original MHD variables and their evolution is not derived directly from the MHD equations~(\ref{MHDequ}-\ref{MHDeqb}). The evolution equations for $\vU$ and $\vB$ are rather postulated a priori, but a number of constraints are imposed on the shell model. In this section, the models are build by imposing on the evolution equations for these vectors as many constraints as possible derived from conservation properties of each of the terms appearing in the original MHD equations.
\newtheorem{prop}{Property}
\begin{prop}\label{P1}: the nonlinear term in the evolution equation for $\vU$ is a sum of two quadratic terms; The first one depends on $\vU$ only and conserves the kinetic energy ${\cal E}^U$ and the kinetic helicity ${\cal H}^k$ 
independently of the value of the field $\vB$; The second term depends on $\vB$ only.
\end{prop}
\begin{prop} \label{P2}: the nonlinear term in the evolution equation for $\vB$ is a sum of two bi-linear terms; The first one must conserve the magnetic energy ${\cal E}^B$ independently of the value of the field $\vU$;
\end{prop}
\begin{prop}\label{P3}: the full nonlinear expression in both the equations for $\vU$ and $\vB$ change sign under the exchange $\vU\leftrightarrow\vB$;
\end{prop}
The dynamical system for the shell vectors can therefore be written:
\begin{align}
d_t \vU & = \vQ(\vU,\vU) - \vQ(\vB,\vB) - \nu\, \vD(\vU)+ \myvector{F}\,, \label{SMMHDgenU}\\
d_t \vB & = \vW(\vU,\vB) - \vW(\vB,\vU) - \eta\, \vD(\vB)\,, \label{SMMHDgenB}
\end{align}
where the term proportional to $\nu$ models the viscous effect, the term proportional to $\eta$ models the Joule effect and $\myvector{F}$ stands for the forcing. The linear operator $\vD$ is defined as follows:
\begin{equation}
\vD(\vX)=(k_1^2\, x_1,k_2^2\, x_2,...,k_N^2\, x_N) \in \mathbb C^N.
\end{equation}

Now, the conservation laws must be enforced. Assuming incompressibiliity, in the ideal limit and in absence of forcing ($\myvector{F}, \nu, \eta \rightarrow 0$), the model is expected to conserve the total energy ${\cal E}^{tot}={\cal E}^U+{\cal E}^B$, the cross helicity ${\cal H}^c$ and the magnetic helicity ${\cal H}^m$.   In terms of shell variables of the model, the energies and the cross helicity are defined for the original MHD equation as
\begin{align} 
{\cal E}^U&= \frac 1 2\ \prodscal{\vU}{\vU}\,, \label{EnU}\\
{\cal E}^B&=\frac 1 2\ \prodscal{\vB}{\vB}\,, \label{EnB} \\
{\cal H}^c &= \prodscal{\vU}{\vB}\,.
\end{align}
The definition of the kinetic helicity and the magnetic helicity requires the expressions for the vorticity $\vO=(o_1,...,o_N)$ and the magnetic potential vector $\vA=(a_1,...,a_N)$. These quantities are not trivially defined in shell models since they require the use of the curl operator. Nevertheless, they should be linear function of the velocity and magnetic field respectively. The kinetic helicity and the magnetic helicity are then defined as follows
\begin{align}
{\cal H}^m &= \prodscal{\vU}{\vO},\\
{\cal H}^k &= \prodscal{\vA}{\vB}.
\end{align}
In terms of conservation laws, the property~\ref{P1} imposes the following constraints that correspond to the conservation of the kinetic energy and the kinetic helicity respectively  by the first quadratic term in the $U$ equation:
\begin{align} 
\prodscal{\vQ(\vU,\vU)}{\vU}=0 \ \forall \vU\ ,\label{P1a}\\
\prodscal{\vQ(\vU,\vU)}{\vO}=0 \ \forall \vU\ .\label{P1b}
\end{align}
Here, the notation ``$\forall \vU$'' must be understood as ``for all possible values of the shell variables $\vU$ as well as $\vO$ that is defined by $\vU$''. The conservation of the magnetic energy by the first quadratic term in the $\vB$ equation (property~\ref{P2}) imposes:
\begin{align} 
\prodscal{\vW(\vU,\vB)}{\vB}=0\ \ \forall \vU,\vB\,.\label{P3a}
\end{align}

The conservations of the total energy and of the cross helicity respectively correspond to
\begin{align} 
\prodscal{\vQ(\vU,\vU)-\vQ(\vB,\vB)}{\vU}&+ \nonumber\\
&\hspace{-25truemm}\prodscal{\vW(\vU,\vB)-\vW(\vB,\vU)}{\vB}=0\ \ \forall \vU,\vB\,,\label{consEtot}\\
\prodscal{\vQ(\vU,\vU)-\vQ(\vB,\vB)}{\vB}&+ \nonumber\\
&\hspace{-25truemm}\prodscal{\vW(\vU,\vB)-\vW(\vB,\vU)}{\vU}=0\ \ \forall \vU,\vB\,.\label{consHcrois}
\end{align}
These two constraints are equivalent since the second is obtained simply from the first under the exchange $(\vU,\vB)\rightarrow (\vB,\vU)$. Hence, the general procedure adopted here shows that in the ideal limit, for a shell model with the structure~(\ref{SMMHDgenU}-\ref{SMMHDgenB}),  the conservation of the total energy ${\cal E}^{tot}$ implies the conservation of the cross helicity ${\cal H}^c$ and vice versa. Moreover, taking into account the constraints~(\ref{P1a}) and~(\ref{P3a}), the conservation of the total energy and cross helicity reduces to:
\begin{align} 
\prodscal{\vQ(\vB,\vB)}{\vU}+ \prodscal{\vW(\vB,\vU)}{\vB}=0\ \ \forall\, \vU,\vB\,,\label{consEtotsimp}
\end{align}
Finally, the conservation of the magnetic helicity imposes the condition:
\begin{align}
\prodscal{\vW(\vU,\vB)}{\vA}+ \prodscal{\vW(\vB,\vU)}{\vA}=0\ \ \forall\, \vU,\vB\,.
\label{consmaghel}
\end{align}
Again, the notation ``$\forall\, \vU,\vB$'' must be understood as ``for all possible values of the shell variables $\vU$ and $\vB$ as well as $\vO$ and $\vA$ that are defined by $\vU$ and $\vB$ respectively''. The specific form of the nonlinear terms in the general shell model~(\ref{SMMHDgenU}-\ref{SMMHDgenB}) can not be defined further without giving explicit definitions for $\vO$ and $\vA$. The choice of the interactions retained in the nonlinear terms (for example: first neighbouring shell or distant shell interactions~\cite{PSshellNonLocal, PSnonLocHLPm}) must also be made explicit in order to reach the final form of the shell model. An example will be treated in Section~\ref{PS}.

If the shell model has to reproduce all the symmetries of the original MHD equation, the following equality could also be imposed:
\begin{equation}
\vW(\vX,\vX)=\vQ(\vX,\vX)\label{nonlineq}
\end{equation}
It is a consequence of the particular way of writing the MHD equations in which all nonlinear terms, including those appearing in the magnetic field equation, are made explicitly divergence free through the application of the projection operation~(\ref{tensorP}). In the example treated in Section~\ref{PS}, this equality appears as a direct consequence of the other constraints imposed on the structure of the shell model. Nevertheless, if the present approach is applied to more complex shell models for MHD, it might be interesting to keep the equality~(\ref{nonlineq}) in mind in order to simplify the nonlinearities as much as possible.


\section {Energy fluxes and energy exchanges \label{TF}}


\subsection{Evolution equations for the shell energies\label{subsec-ener}}

The kinetic and magnetic energies associated with the shell $s_n$ are defined as $e^u_n=\prodscal{\vU_n}{\vU_n}/2$ and $e^b_n=\prodscal{\vB_n}{\vB_n}/2$. The evolution equations for these quantities are easily obtained in the inviscid and unforced limit:
\begin{align}
d_t e^u_n &= T^u_n= \prodscal{\vQ(\vU,\vU)-\vQ(\vB,\vB)}{\vU_n}\,, \label{EuEvolve} \\
d_t e^b_n &= T^b_n= \prodscal{\vW(\vU,\vB)-\vW(\vB,\vU)}{\vB_n}\,. \label{EbEvolve}
\end{align}
The quantity $T^u_n$ corresponds to the energy transferred into the velocity field in shell $s_n$ and coming from either the velocity or the magnetic fields. Since the first term of equation~(\ref{EuEvolve}) conserves the total kinetic energy (cf.~(\ref{P1a})), it is identified as the rate of energy $T^{uu}_n$ flowing from the complete velocity field into the velocity field in the $n$-th shell.  The second term of (\ref{EuEvolve}) must then account for the energy coming from the magnetic field ($T^{ub}_n$), i.e.,
\begin{align}
T^{uu}_n&=\prodscal{\vQ(\vU,\vU)}{\vU_n}\,,\label{TuuDef}\\
T^{ub}_n&=-\prodscal{\vQ(\vB,\vB)}{\vU_n}\,.\label{TubDef}
\end{align}
Similarly, $T^b_n$ corresponds to the energy transferred into the magnetic field in shell $s_n$ and coming from either the velocity field or the magnetic field. The first term of equation~(\ref{EbEvolve}) conserves the total magnetic energy (cf. (\ref{P3a})), and is identified with the rate of energy flowing from the complete magnetic field to the magnetic field of the $n$-th shell.  The second term of Eq.~(\ref{EbEvolve}) corresponds to the energy flowing to the $B_n$ shell from the complete velocity field, i.e.,
\begin{align}
T^{bb}_n&=\prodscal{\vW(\vU,\vB)}{\vB_n}\,,\\
T^{bu}_n&=-\prodscal{\vW(\vB,\vU)}{\vB_n}\,.\label{TbuDef}
\end{align}
With this notation, the evolution equations for $e^u_n$ and $e^b_n$ become (with dissipative and forcing terms):
\begin{align} 
{d_t e^u_n} &= T^{uu}_n+T^{ub}_n- 2\ \nu\  k_n^2\ e^u_n+ P^f_n\,, \label{eqeun} \\
{d_t e^b_n} &= T^{bb}_n+T^{bu}_n- 2\ \eta\ k_n^2\ e^b_n\,, \label{eqebn}
\end{align}
where $P^f_n=\prodscal{\myvector{F}}{\vU_n}$ is the kinetic energy injection rate into the shell $s_n$ due to the external forcing.

 It is also convenient to introduce the following decomposition of the vectors of shell variables:
\begin{align}
\vX^<_i&=(x_1,x_2,...,x_{i-1},x_i,0,...,0) \in \mathbb C^N \,,\\
\vX^>_i&=(0,0,...,0,x_{i+1},x_{i+2},...,x_N) \in \mathbb C^N \,,\\
\vX&=\vX^<_i+\vX^>_i \,.
\end{align}
 where $i$ can take any value between 1 and $N$.
The kinetic energy contained in the vector $\vU_n^<$ is simply given by $E^{U^<}_n=\prodscal{\vU_n^<}{\vU_n^<}/2=\sum_{j=1}^{n} e^u_j$. The magnetic energy contained in the vector $\vB_n^<$ is defined similarly. The evolution of these quantities are easily derived from  the relation~(\ref{eqeun}-\ref{eqebn}) :
\begin{align}
d_t E^{U^<}_n &= \sum_{j=1}^{n}  T^{uu}_j+\sum_{j=1}^{n} T^{ub}_j - D_{\nu n}^< + P^{f<}_n\,, \label{eqEu<} \\
d_t E^{B^<}_n &= \sum_{j=1}^{n}  T^{bb}_j+\sum_{j=1}^{n} T^{bu}_j - D_{\eta n}^< \,, \label{eqEb<}
\end{align}
where $P^{f<}_n=\prodscal{\myvector{F}}{\vU_n^<}$ is the injection rate of energy in $\vU^<_n$ due to the forcing, and $D_{\nu n}^<=\nu\,\prodscal{\vD(\vU)}{\vU_n^<}$ and $D_{\eta n}^<=\eta\,\prodscal{\vD(\vB)}{\vB_n^<}$ are the dissipative terms for $U_n^<$ and $B_n^<$ respectively.


\subsection{Energy Fluxes\label{subsec-fluxes}}

The nonlinear terms in the equations~(\ref{eqEu<}-\ref{eqEb<}) correspond to the nonlinear energy fluxes that enter or leave the sphere of radius $k_0 \lambda^n$. These fluxes can be further specified. Indeed, the first sum in the right hand side of the equation~(\ref{eqEu<}) comes from the quadratic $\vQ(\vU,\vU)$ term which conserves the total kinetic energy. Hence, this first sum must correspond to the kinetic energy flux $\Pi^{U>}_{U<}(n)$ from $\vU^>_n$ to $\vU^<_n$:
\begin{equation}
\Pi^{U>}_{U<} (n)=\sum_{j=1}^{n} T^{uu}_j=\prodscal{\vQ(\vU,\vU)}{\vU^<_n}\,.\label{fluxuu}
\end{equation}
The anti-symmetry property for the fluxes can be used to define the opposite transfer: $\Pi^{U<}_{U>} (n) = - \Pi^{U>}_{U<} (n)$. It simply expresses that the energy gained by $\vU^<_i$ due to the nonlinear interaction is equal and opposite to the energy lost by $\vU^>_i$. The magnetic energy fluxes can be similarly defined as
\begin{equation}
\Pi^{B>}_{B<} (n)=\sum_{j=1}^{n} T^{bb}_j=\prodscal{\vW(\vU,\vB)}{\vB^<_n}\,.\label{fluxbb}
\end{equation}
The ``cross'' fluxes between the velocity and the magnetic field can also be defined systematically. The second sum in the right hand side of equation~(\ref{eqEb<}) corresponds to the flux of energy from $\vB_n^<$ to $\vU$ and readily leads to the following definitions:
\begin{align}
\Pi^{U}_{B<}(n)&= \sum_{j=1}^{n} T^{bu}_j= - \prodscal{\vW(\vB,\vU)}{\vB^<_n}\,, \label{fluxub1}\\
\Pi^{U}_{B>}(n)&= \sum_{j=n+1}^{N} T^{bu}_j= \prodscal{\vW(\vB,\vU)}{\vB^>_n}\,.\label{fluxub2}
\end{align}
Since these terms are linear in $\vU$, each of them can easily be split into two contributions related to $\vU_n^<$ and $\vU_n^>$ respectively:  
\begin{align}
\Pi^{U<}_{B<}(n)&=-\Pi^{B<}_{U<}(n)=-\prodscal{\vW(\vB,\vU^<_n)}{\vB^<_n}\,,\label{fluxub3}\\
\Pi^{U>}_{B<}(n)&=-\Pi^{B<}_{U>}(n)=-\prodscal{\vW(\vB,\vU^>_n)}{\vB^<_n}\,,\label{fluxub4}\\
\Pi^{U<}_{B>}(n)&=-\Pi^{B>}_{U<}(n)=-\prodscal{\vW(\vB,\vU^<_n)}{\vB^>_n}\,,\label{fluxub5}\\
\Pi^{U>}_{B>}(n)&=-\Pi^{B>}_{U>}(n)=-\prodscal{\vW(\vB,\vU^>_n)}{\vB^>_n}\,.\label{fluxub6}
\end{align}
The formula~(\ref{fluxuu}-\ref{fluxub6}) shows that the various fluxes can be defined univocally, almost independently of the structure of the shell model as long as the terms conserving kinetic and magnetic energy have been identified. It must be stressed that, at this stage, the exact expressions for the nonlinear terms $\vQ$ and $\vW$ are not needed.


\subsection{Shell-to-shell energy exchanges\label{subsec-transfers}}

The expression for some of the energy exchanges between two shells may be derived from the above analysis. For instance, the quantity $T^{bu}_n$ has been identified as the energy flux from the entire velocity field to the magnetic field associated to the shell $s_n$. The expansion~(\ref{vecexpan}) for $\vU$ can be inserted into the term $T^{bu}_n$ and leads to:
\begin{equation}
T^{bu}_n= \sum_{m=1}^N -\prodscal{\vW(\vB,\vU_m)}{\vB_n}=\sum_{m=1}^N T^{bu}_{nm} \,,\label{propTbu}
\end{equation}
where each term in this sum can now be identified as the shell-to-shell energy exchange rate from the velocity field in the shell $s_m$ to the magnetic field in the shell $s_n$:
\begin{equation}
T^{bu}_{nm}=-\prodscal{\vW(\vB,\vU_m)}{\vB_n}\,.\label{Tbunm}
\end{equation}
Similarly, by inserting the expansion~(\ref{vecexpan}) for $\vB$ into the term $T^{bb}_n$, it is possible to identify the shell-to-shell energy exchange rate from the magnetic field in the shell $s_m$ to the magnetic field in the shell $s_n$ as follows:
\begin{equation}
T^{bb}_{mn}=\prodscal{\vW(\vU,\vB_m)}{\vB_n}\,,\label{Tbbnm}
\end{equation}
Since the quantities $T^{xy}_{nm}$ are a shell-to-shell energy exchange rate (the notation $xy$ is referred to as general exchange and it can take values $uu$, $ub$, $bu$ or $bb$), the following anti-symmetry property is to be satisfied:
\begin{equation} 
T^{xy}_{mn} = -T^{yx}_{nm}. \label{anti-sym}
\end{equation}

It is worth mentioning that the present analysis does not lead to a simple definition of the shell-to-shell kinetic energy exchanges $T^{uu}_{nm}$. This is due to the presence of three velocity variables in the expression for the $\vU$-to-$\vU$ transfers that prevents a simple identification of the origin of the kinetic energy flux. Nevertheless, considering the relation~(\ref{nonlineq}), the quantity $T^{uu}_n$~(\ref{TuuDef}) can be rewritten as follows:
\begin{equation} 
T^{uu}_{n} = \prodscal{\vW(\vU,\vU)}{\vU_n}\,,
\end{equation}
and, by analogy with the expression~(\ref{Tbbnm}), it is reasonable to adopt the following definition:
\begin{equation} 
T^{uu}_{nm} = \prodscal{\vW(\vU,\vU_m)}{\vU_n}\,.\label{Tuunm}
\end{equation}
The shell-to-shell energy exchanges give a more refined picture of the dynamics in the shell model than the fluxes. It is thus expected that these fluxes can be reconstructed from all the $T^{xy}_{nm}$. The general formula are given by:
\begin{align}
\Pi^{Y<}_{X>}(n)& = \sum_{i=n+1}^N \sum_{j=1}^n T^{xy}_{ij}\,,\\
\Pi^{Y<}_{X<}(n)& = \sum_{i=1}^{n} \sum_{j=1}^n T^{xy}_{ij}\,,\\
\Pi^{Y>}_{X>}(n)& = \sum_{i=n+1}^N \sum_{j=n+1}^N T^{xy}_{ij}\,.
\end{align}
As a direct consequence of the property~(\ref{anti-sym}), the same anti-symmetry property holds for the energy fluxes. 

In the next section we will focus on a specific model adopted by Stepanov and Plunian~\cite{PS1}. We will derive the formulas for the energy fluxes and compute them numerically.


\section{Study of a GOY shell model for MHD turbulence \label{PS}}


\subsection{Derivation of the shell model}

The results derived in Sections~\ref{GenShel} and~\ref{TF} are valid for any shell model for MHD that use only one complex number per shell for each field (velocity and magnetic) and for which the properties~(\textbf{1-3}) are satisfied. As long as the vorticity and the magnetic potential vector have not been defined explicitly, it is not possible to specify further the exact structure of the shell model, i.e. the structure of the nonlinear terms $\vQ$ and $\vW$. In this section, we revisit the GOY-like shell model for MHD turbulence studied by Stepanov and Plunian~\cite{PS1}, and apply  the formalism discussed in Sections~\ref{GenShel} and~\ref{TF} to this model. The shell model is defined by the following expressions for the nonlinear  $Q$ and $W$ terms:
\begin{align} 
q_n(X,X)&=i k_n (\alpha_1 x_{n+1}^* x_{n+2}^* + \alpha_2 x_{n-1}^* x_{n+1}^*\nonumber\\
&\hspace{8truemm} + \alpha_3 x_{n-2}^* x_{n-1}^*)\,,  \label{QDef} \\
w_n(X,Y)&= ik_n(\beta_1 x_{n+1}^* y_{n+2}^* + \beta_2 x_{n-1}^* y_{n+1}^* \nonumber \\ 
&\hspace{8truemm}+ \beta_3 x_{n-2}^* y_{n-1}^* + \beta_4 y_{n+1}^* x_{n+2}^* \nonumber \\ 
&\hspace{12truemm} +\beta_5 y_{n-1}^* x_{n+1}^* +  \beta_6 y_{n-2}^* x_{n-1}^* )\,.\label{WDef}
\end{align}
This shell model is fully determined if the following definitions for the vorticity and the magnetic potential vector are also adopted:
\begin{align}
o_i &= (-1)^i  u_i k_i \,,\\
a_i &= (-1)^i  b_i / k_i \,.
\end{align}
Imposing the conditions derived in the previous section from the various conservation laws (\ref{P1a}, \ref{P1b}, \ref{P3a}, \ref{consEtotsimp} and \ref{consmaghel}) lead to the following values of the parameters $\alpha_i$ and $\beta_i$:
\begin{align}
\alpha_2&= -\alpha_1 \frac {\lambda -1} {\lambda^2}&
\alpha_3&= -\alpha_1 \frac 1 {\lambda^3}\nonumber \\
\beta_1&= \alpha_1 \frac {\lambda^2 + \lambda + 1} {2\lambda (\lambda+1)}&
\beta_2&= -\alpha_1 \frac {\lambda^2 - \lambda - 1} {2\lambda^2 (\lambda+1)}\nonumber \\
\beta_3&= \alpha_1 \frac {\lambda^2 - \lambda - 1} {2\lambda^3  (\lambda+1)}&
\beta_4&= \alpha_1 \frac {\lambda^2 + \lambda - 1} {2\lambda (\lambda+1)}\nonumber \\
\beta_5&= -\alpha_1 \frac {\lambda^2 + \lambda - 1} {2\lambda^2 (\lambda+1)}&
\beta_6&= -\alpha_1 \frac {\lambda^2 + \lambda + 1} {2\lambda^3 (\lambda+1)}\nonumber 
\end{align}
As discussed at the end of Section \ref{GenShel}, this shell model also satisfies the constraint~(\ref{nonlineq}). It is indeed easy to verify that these parameters satisfy the following equalities: $\beta_1+\beta_4=\alpha_1$, $\beta_2+\beta_5=\alpha_2$ and $\beta_3+\beta_6=\alpha_3$. 

In order to verify that the model derived here is exactly the same as the model discussed in~\cite{PS1}, the dynamical system (\ref{SMMHDgenU}-\ref{SMMHDgenB}) can then be rewritten after a few algebraic manipulations as
\begin{align} 
{d_t u_n} & = ik_n \Big(p_n(\vU,\vU) - p_n(\vB,\vB)\Big)-\nu  k_n^2 u_n+ f_n \label{ModPSu}\\
{d_t b_n} & = ik_n \Big(v_n(\vU,\vB) - v_n(\vB,\vU)\Big)-\eta k_n^2 b_n \label{ModPSb}
\end{align} 
where
\begin{align}
p_n(\vX,\vX)&=\alpha_1\ \Big(x_{n+1}^* x_{n+2}^* - \frac{\lambda-1}{\lambda^2}\ x_{n-1}^* x_{n+1}^*\nonumber\\
&\hspace{25truemm}-\frac{1}{\lambda^3}\ x_{n-2}^* x_{n-1}^*\Big)\,,\label{pn}\\
v_n(\vX,\vY)&=\frac{\alpha_1}{\lambda (\lambda+1)}\ \Big(x_{n+1}^* y_{n+2}^* + x_{n-1}^* y_{n+1}^* \nonumber\\
&\hspace{25truemm}+x_{n-2}^* y_{n-1}^*\Big) \,. \label{vn}
\end{align}
With these coefficients, the model~(\ref{ModPSu}-\ref{ModPSb}) is clearly the same as the one derived by Plunian and Stepanov~\cite{PS1}. Our interpretation of some of the shell-to-shell energy exchanges and the energy fluxes derived in the Section~\ref{TF} and computed in the next section differs from those of Stepanov and Plunian~\cite{PS1}.  When we compare the two approaches carefully, we find that the velocity to velocity energy flux $\Pi^{U<}_{U>}$ and the total fluxes are the same for both the formalism, but other fluxes involving the magnetic field are different.   This is due to the fact that the complete function $\vW$ is never computed in~\cite{PS1} because the property~\ref{P2} was not used explicitly in the derivation of the shell model. In particular, the part of the bilinear term that conserves the magnetic energy in the magnetic field equation was not identified. It was not needed to derive completely the model coefficient. However, this identification is  needed if the energy fluxes have to be defined unambiguously, which is the main objective of this work, but not of the approach developed by Plunian and Stepanov.


\subsection{Numerical results \label{res}}

In order to compute the energy fluxes, we simulate the shell model~(\ref{ModPSu}-\ref{ModPSb}) with $\nu =10^{-9}$ and $\eta =10^{-6}$. The magnetic Prandtl number is then $P_M=\nu/\eta=10^{-3}$. The shells ratio is taken to be the golden mean: $\lambda =(1+\sqrt 5)/2$. We take the number of shells as $N=36$ and apply non-helical forcing to $s_4$, $s_5$ and $s_6$ according to the scheme prescribed by Stepanov and Plunian~\cite{PS1}. The energy injection rate $\epsilon$ is 1.   We evolve the shell model till our system reaches a steady-state, and then we  compute the energy fluxes by averaging over many time frames.  Up to slight differences due to the time steps or the initial conditions,  several steady state  results obtained from our simulation very well reproduce the results presented by Plunian and Stepanov~\cite{PS1}. 

\begin{figure}[ht]
\begin{center}
\includegraphics[width=8cm]{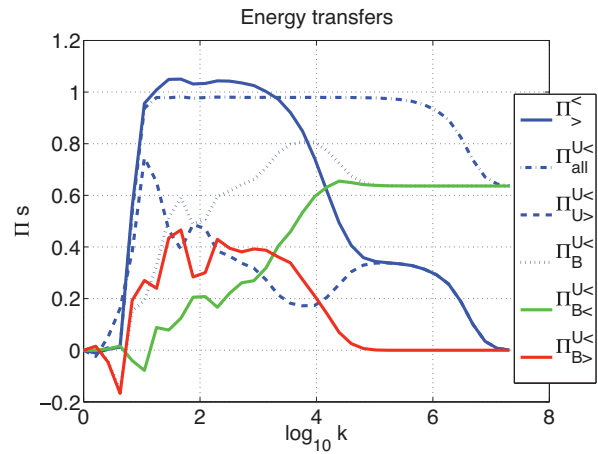}
\caption{Energy fluxes in function of the logarithm of $k_n$. $\nu=10^{-9}$ and $P_M=10^{-3}$.}\label{PSFlu}
\end{center}
\end{figure}

In Fig.~\ref{PSFlu} we present some of the energy fluxes as a function of wavenumber.  We present the energy fluxes $\Pi^{U<}_{U>}(n)$, $\Pi^{U<}_{B>}(n)$ and $\Pi^{U<}_{B<}(n)$ that are the rate of energy leaving a wavenumber sphere of radius $k_n$.  In addition we also report the energy fluxes  $\Pi^{U<}_{B<}$   and $\Pi^{U<}_{all}$ that are defined as
\begin{align}
\Pi^{U<}_B  &=\Pi^{U<}_{B<}+\Pi^{U<}_{B>}\,,\label{flux-umb}\\
\Pi^{U<}_{\textrm{all}} &=\Pi^{U<}_{U>}+\Pi^{U<}_{B<}+\Pi^{U<}_{B>}=\Pi^{U<}_B+\Pi^{U<}_{U>}\,,\label{flux-umtot}
\end{align}
Finally, the total energy flux $\Pi^{<}_> $  leaving the sphere of wave vector smaller than $k_n$, from either the velocity or the magnetic field, is also presented on Figure~\ref{PSFlu}:
\begin{equation}
\Pi^{<}_> =\Pi^{U<}_{U>}+\Pi^{U<}_{B>}+\Pi^{B<}_{U>}+\Pi^{B<}_{B>}\,,\label{flux-ubtot}
\end{equation}
Our energy fluxes $\Pi^{<}_> $ and $\Pi^{U<}_{U>}$ are in good agreement with the corresponding fluxes reported by Stepanov and Plunian~\cite{PS1}, which is expected since these fluxes are computed by equivalent schemes in both these work. However the energy fluxes from the velocity field to the magnetic field and vice versa do not match because these fluxes are computed differently in these schemes.  Our flux formulas have the advantage of being defined by a more systematic method compared to that of Stepanov and Plunian~\cite{PS1}.

\begin{figure}[ht]
\begin{center}
\includegraphics[width=8cm]{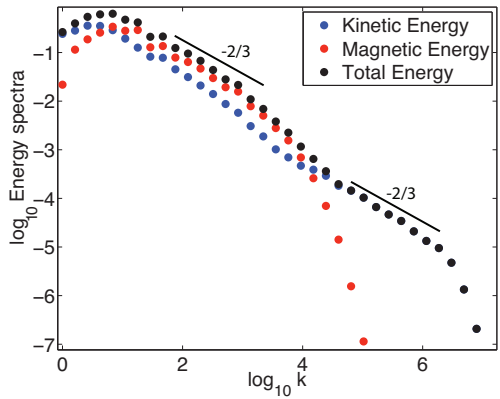}
\caption{Kinetic and magnetic energy spectra $\nu=10^{-9}$ and $P_M=10^{-3}$.}\label{EnSpec}
\end{center}
\end{figure}

We also compute kinetic, magnetic, and total energy spectra in the steady-state.  We observe that till $k\sim 10^4$ both the kinetic and magnetic energy show power law behaviour with -2/3 spectral exponent consistent with Kolmogorov's spectrum (Fig.~\ref{EnSpec}).   After $k\sim 10^4$, the magnetic energy decays exponentially due to the Joule dissipation, while the kinetic energy continues to exhibit power law behaviour with the same spectral exponent of -2/3.  Stepanov and Plunian~\cite{PS1} report two different spectral regimes at steady state.  They report -1 spectral exponent at lower wavenumbers  for both velocity and magnetic field.  For higher wavenumbers, they report exponential decay for the magnetic energy and -2/3 spectral index for the kinetic energy.  Our result differs from that of Stepanov and Plunian at lower waveumbers.  We believe this discrepancy is possibly due to a short range of wavenumbers that makes the determination of the exponent quite difficult.

To understand the two spectral regimes better we focus on the energy fluxes in these regimes. In the first regime (the lower wavenumber), the energy flux $\Pi^{U<}_{\textrm{all}}$ and $\Pi^{<}_>$ are approximately equal. The kinetic and magnetic energy spectra show that both the velocity and the magnetic field are active in this regime.  Hence it is reasonable to expect Kolmogorov-like energy spectrum for MHD turbulence according to the existing MHD turbulence phenomenologies (for review, see \cite{ver04}). In the second regime (the larger wavenumbers), the magnetic field is damped heavily by Joule dissipation, while the velocity field  evolves according to  the nonlinear term  $\vQ(\vU,\vU)$ which corresponds to the convective nonlinearity $\bu\cdot\nabla\, \bu$.  Hence the turbulence here is almost purely hydrodynamics except that $U^<$ shells still supply energy flux to the magnetic shells of the dissipative regime (small amplitudes).  Therefore, it is reasonable to expect Kolmogorov's energy spectrum for the velocity field in this regime as well but with a smaller energy flux.  Hence, our energy spectra and energy fluxes shown in Figs.~\ref{PSFlu} and~\ref{EnSpec} are consistent since they exhibit almost the same slope but different amplitudes.

In the following section we summarize our results.


\section{Conclusion}

A general derivation of shell models for MHD has been proposed. The conservation of the traditional ideal invariants of three-dimensional MHD turbulence is expressed as general constraints that must be satisfied by the nonlinear terms in the shell model. The conservation of the kinetic helicity and kinetic energy by the hydrodynamic shell model in absence of magnetic field also leads to constraints on the nonlinearities. The similarity between the original MHD equations and the shell model is pushed one step further by identifying one term in the magnetic field equation in the shell model that conserves the magnetic energy. It corresponds to the advection of magnetic field by the velocity in the MHD equations. This procedure is presented using a very general formalism which leads to a number of interesting results.

We show that the conservation of the cross helicity and the conservation of the total energy are equivalent in shell models. This equivalence is a direct consequence of the symmetries of the MHD equations expressed by the general properties 1-3 presented in Section~\ref{GenShel}.

The expressions for the energy fluxes that are valid independently of the specific structure of the nonlinear couplings between the shell variables have been derived. The knowledge of these fluxes is quite important when the shell models are used to explore dynamo regime. These expressions could even be used to derive shell models that would maximise or minimise certain energy transfers depending on the physics that has to be modelled. 

Also, expressions for the shell-to-shell energy exchanges are derived. Like in the original MHD equations, the energy exchange mechanisms in shell models unavoidably involve three degrees of freedom (triadic interaction)~\cite{doro90,Waleffe,davees01,ver04, deveca05,cadekn06,almipo05b,mialpo05}. It is thus not obvious to derive expression for shell-to-shell energy exchanges that are viewed as energy transfers between only two degrees of freedom. Nevertheless, the formalism presented in Section~\ref{GenShel} yields a very natural identification of most of these energy exchanges. The only exception concerns the $\vU$-to-$\vU$ energy exchanges. A simple expression is however also proposed for these quantities by analogy with the $\vB$-to-$\vB$ energy exchanges.

Another property of the formalism presented here is the clear separation between the treatment of the conservation law and the assumptions that have to be made to define both the magnetic and for the kinetic helicities. Because these helicities involve quantities that are defined using the curl operator, they are not very well adapted to shell models. It is thus quite appropriate to clearly present the expressions for the vorticity and the magnetic potential as additional assumptions required to fully specify the structure of the shell model.
 
The procedure has been applied to a specific class of shell models based on first neighbour couplings, know as the GOY model. It has been shown that the general constraints naturally leads to the already derived GOY-MHD shell model~\cite{PS1}. However, the interpretation of the energy fluxes appears to be simpler in the present formalism. 

Several extensions to this work could be considered. Shell models using distant interactions between the shell variables~\cite{PSnonLocHLPm} could be analysed using the same formalism. Also, despite the fact that the presentation has been made for shell models with one complex number per shell and per field (velocity and magnetic), extending the present formalism to shell models with more degrees of freedom should be quite obvious. Finally, it would be interesting to explore other shell models based on alternative definitions for both the vorticity and the magnetic potential.


\section*{Acknowledgements}

The authors are pleased to acknowledge very fruitful discussions with F. Plunian and R. Stepanov during the 2007 MHD summer programme at ULB (Brussels, Belgium). This work has been supported by the contract of association EURATOM - Belgian state. The content of the publication is the sole responsibility of the authors and it does not necessarily represent the views of the Commission or its services. D.C. and T.L. are supported by the Fonds de la Recherche Scientifique (Belgium).  MKV thanks the Physique Statistique et Plasmas group at the University Libre de Bruxelles for the kind hospitality and financial support during his long leave when this work was undertaken.


\begin{thebibliography}{10}

\bibitem{Moffat}
H.~K. Moffat.
\newblock {\em Magnetic Fields Generation in Electrically Conducting Fluids}.
\newblock Cambridge University Press, 1978.

\bibitem{Krause}
F.~Krause and K.H. Radler.
\newblock {\em Mean-Field Magnetohydrodynamics and Dynamo Theory, Cambdrige}.
\newblock Pergamon Press, Oxford, 1980.

\bibitem{Bran}
A.~Brandenburg and K.~Subramanian.
\newblock {\em Phys. Rep.}, 417(1), 2005.

\bibitem{Monch}
R.~Monchaux, M.~Berhanu, M.~Moulin, and P.~Odier.
\newblock {\em Phys. Rev. Lett.}, 98, 2007.

\bibitem{Fauv}
M.~Berhanu, R.~Monchaux, S.~Fauve, and N.~Mordant.
\newblock {\em Europhys. Lett.}, 77, 2007.

\bibitem{bif03}
L.~Biferale.
\newblock {\em Ann. Rev. Fluid Mech.}, 35:441--468, 2003.

\bibitem{Gletzer}
E.B. Gletzer.
\newblock {\em Soviet Phys., Doklady}, 18(216), 1973.

\bibitem{YamadaOhkitani}
M.~Yamada and K.~Ohkitani.
\newblock {\em Journal of the Physical Society of Japan}, 56, 1987.

\bibitem{Lrt}
C.~Gloaguen, J.~L\'eorat, A.~Pouquet, and R.~Grappin.
\newblock {\em Physica D}, 17:154--182, 1985.

\bibitem{Frick}
P.~Frick and S.~Sokoloff.
\newblock {\em Phys. Rev. E}, 57(4), 1998.

\bibitem{Gilbert}
T.Gilbert and D.~Mitra.
\newblock {\em {Phys. Rev. E}}, {69}({5, Part 2}), {MAY} {2004}.

\bibitem{PS1}
R. Stepanov and F. Plunian.
\newblock {\em Journal of Turbulence}, 7(39), 2006.


\bibitem{davees01}
G. Dar, M.K. Verma and V. Eswaran.
\newblock {\em Physica D}, 3:207--225, 2001.

\bibitem{ver04}
M. K. Verma,
\newblock {\em Phys. Report}, 401: 229--380, 2004.

\bibitem{Sabra}
V.~S. L'vov, E.Podivilov, A.~Pomyalov, I.~Procaccia, and D.~Vandembroucq.
\newblock {\em Phys. Rev. E}, 58(2), 1998.

\bibitem{PSshellNonLocal}
F.~Plunian and R.~Stepanov.
\newblock {\em New Journal of Physics}, 9(8):294, 2007.

\bibitem{PSnonLocHLPm}
R.~Stepanov and F.~Plunian.
\newblock {\em The Astrophysical Journal}, 680(1):809--815, 2008.

\bibitem{doro90}
J.A. Domaradski and R.S. Rogallo.
\newblock {\em Phys. Fluids A}, 2:413--426, 1990.

\bibitem{Waleffe}
F.~Waleffe.
\newblock {\em Phys. Fluids A}, 4(2), 1992.

\bibitem{deveca05}
O. Debliquy, M. K. Verma and D. Carati,
\newblock {\em Phys. Plasmas}, 12: 042309, 2005.

\bibitem{cadekn06}
D. Carati, O. Debliquy, B. Knaepen, B. Teaca and M. Verma.
\newblock {\em Journal of Turbulence}, 3:207--225, 2006.

\bibitem{almipo05b}
A.~Alexakis, P.D. Mininni, and A.~Pouquet.
\newblock {\em Phys. Rev. E}, 72:046301, 2005.

\bibitem{mialpo05}
P.D. Mininni, A.~Alexakis and A.~Pouquet.
\newblock {\em Phys. Rev. E}, 72:046302, 2005.

\end{thebibliography}
\end{document}